\documentclass[aip,jcp,reprint,noshowkeys,superscriptaddress]{revtex4-1}
\usepackage{graphicx,dcolumn,bm,xcolor,microtype,multirow,amscd,amsmath,amssymb,amsfonts,physics,longtable,wrapfig,bbold,siunitx,xspace}
\usepackage[version=4]{mhchem}

\usepackage{mathtools}

\usepackage[utf8]{inputenc}
\usepackage[T1]{fontenc}

\usepackage{hyperref}
\hypersetup{
    colorlinks,
    linkcolor={red!50!black},
    citecolor={red!70!black},
    urlcolor={red!80!black}
}

\usepackage{listings}
\definecolor{codegreen}{rgb}{0.58,0.4,0.2}
\definecolor{codegray}{rgb}{0.5,0.5,0.5}
\definecolor{codepurple}{rgb}{0.25,0.35,0.55}
\definecolor{codeblue}{rgb}{0.30,0.60,0.8}
\definecolor{backcolour}{rgb}{0.98,0.98,0.98}
\definecolor{mygray}{rgb}{0.5,0.5,0.5}

\definecolor{sqred}{rgb}{0.85,0.1,0.1}
\definecolor{sqgreen}{rgb}{0.25,0.65,0.15}
\definecolor{sqorange}{rgb}{0.90,0.50,0.15}
\definecolor{sqblue}{rgb}{0.10,0.3,0.60}

\lstdefinestyle{mystyle}{
    backgroundcolor=\color{backcolour},
    commentstyle=\color{codegreen},
    keywordstyle=\color{codeblue},
    numberstyle=\tiny\color{codegray},
    stringstyle=\color{codepurple},
    basicstyle=\ttfamily\footnotesize,
    breakatwhitespace=false,
    breaklines=true,
    captionpos=b,
    keepspaces=true,
    numbers=left,
    numbersep=5pt,
    numberstyle=\ttfamily\tiny\color{mygray},
    showspaces=false,
    showstringspaces=false,
    showtabs=false,
    tabsize=2
  }

  \usepackage[]{notes2bib}
  \usepackage{siunitx}

  \newcolumntype{d}{D{.}{.}{-1}}

\newcommand*{\ron}{n}

\def\md{\mathrm{d}}
\def\ud{\mathrm{d}}

\def\nm1{{N-1}}
\def\kin{\text{kin}}
\def\cond{\text{cond}}
\def\rr{{\bm{r}}}
\def\sce{\text{SCE}}

\DeclareUnicodeCharacter{2212}{-}

\newcommand{\ICP}{Universit\'e Paris-Saclay, CNRS, Institut de Chimie Physique UMR8000, 91405, Orsay, France}

\begin{document}	

\title{Modelling the Kohn-Sham Potential for Molecular Dissociation with Orbital-independent Functionals: a Proof of Principle}

\author{Sara \surname{Giarrusso}}
	\email{sara.giarrusso@universite-paris-saclay.fr}
	\affiliation{\ICP}
	
\author{Federica \surname{Agostini}}
	\affiliation{\ICP}

\begin{abstract}
We model the Hartree-exchange-correlation potential of Kohn-Sham density-functional theory adopting a novel strategy inspired by the strictly-correlated-electrons limit and relying on the exact decomposition of the potential based on the exact factorization formalism.
Starting with accurate density and conditional potential for a one-dimensional model of a stretched heteronuclear molecule, we provide a proof-of-principle example of an approximation that accurately reproduces the step of the exact Kohn-Sham potential without resorting to the Kohn-Sham orbitals, virtual or occupied. 
We also test our strategy using the approximate conditional potentials corresponding to the strictly-correlated-electron and the exact-exchange functionals. The results are encouraging in that the initial approximations are modified in the qualitatively correct way: decreasing correlation in the former and increasing it in the latter. 
\end{abstract}

\maketitle

Kohn-Sham (KS) density functional theory (DFT) is an elegant and computationally-efficient theory for electronic structure whose great success hinges on the mapping of a physical system of interacting electrons onto an auxiliary, formally non-interacting one, i.e. the KS system.~\cite{HohKoh-PR-64, KohSha-PR-65} 
The KS external potential comprises the external potential of the original system and the Hartree-exchange-correlation (Hartree-XC) potential, whose XC part has to be approximated. Accurately capturing chemical and physical properties relies on the quality of the approximations of the XC potential. Nonetheless, fundamental processes in chemistry, such as bond breaking or charge localization, or properties, such as electric polarizabilities and band gaps, can be captured correctly only if the XC potential manifests \textsl{non-intuitive} features, such as ``peaks'' and ``steps''.~\cite{AlmBar-INC-85,BuiBaeSni-PRA-89,GriLeeBae-JCP-94, GriBae-PRA-96, GriLeeBae-JCP-96,GriBae-TCA-97, BaeGri-JPCA-97,GisSchGriBaeSniChaKir-PRL-99, TemMarMai-JCTC-09, HelTokRub-JCP-09,RyaKohSta-PRL-15, HodRamGod-PRB-16, YinBroLopVarGorLor-PRB-16, RyaOspSta-JCP-17, HodKraSchGro-JPCL-17,KaiKum-PRA-18,GiaVucGor-JCTC-18, GiaGorGie-EPJB-18, GiaGor-JPCA-20, GiaNeuBaeGie-JCTC-22, KocKraSch-JPCL-21,KocKraSch-PRR-23,DarLacMai-CPR-22} 

Peaks and steps are necessary to enforce all the correlation effects of the original system in a non-interacting system. For instance, without the step structure around the more electronegative atom in the KS potential for the stretched LiH molecule, there would be a spurious accumulation of electron density on the hydrogen atom (more electronegative), resulting in fractionally charged fragments. Such accumulation of charge is impeded by electronic repulsion in the interacting system. 
Conversely, the step structure around H in the exact KS potential \textit{de facto} aligns the ionization potentials of the two fragments in the non-interacting system, preventing spurious electron density accumulation and, therefore, unphysical fractional charges.~\cite{AlmBar-INC-85, GriBae-PRA-96}

Mathematically, peaks and steps emerge as a consequence of the non-local dependence of the KS potential on the electronic density. It is logical, then, that the numerically-efficient local or semilocal approximate XC functionals fail in capturing these features. On the other hand, many orbital-dependent functionals require virtual KS orbitals to incorporate non locality,~\cite{Bae-PRL-01,GriBae-IJQC-06,Fur-PRB-01,FucGon-PRB-02,GruGriBae-JCP-03,Cha-JCP-12,HelGou-JCTC-19,WanBae-PRL-22} which makes them expensive thus limiting their applicability.

In this work, we propose a novel strategy aiming to model correctly the step structure of the exact potential in the molecular dissociation limit. We work within an orbital-independent theoretical framework, which allows us to approximate accurately the Hartree-XC potential based on a simple numerical procedure. The first key ingredient of this strategy is a decomposition of the Hartree-XC potential able to isolate the contributions responsible for peaks and steps. It was introduced by Buijse \textit{et al.}~\cite{BuiBaeSni-PRA-89} and is based on the theory of the \textsl{conditional} amplitude, also known as exact electron factorization (EEF).~\cite{Hun-IJQC-75-1, Hun-IJQC-75-2, GiaVucGor-JCTC-18, GiaGor-JPCA-20,KocKraSch-JPCL-21,KocKraSch-PRR-23, GiaGorAgo-CPC-24} Our idea, which is inspired by the mathematical structure of the strong-interaction or strictly-correlated-electrons (SCE) limit of DFT,~\cite{Sei-PRA-99, SeiGorSav-PRA-07,GorVigSei-JCTC-09, MirSeiGor-JCTC-12} is to approximate the components of the Hartree-XC potential using their expressions in terms of the second key ingredient of the strategy, namely the \textsl{conditional} density. While we test our strategy on a model stretched heteronuclear molecule for which the \textsl{conditional} density can be calculated very accurately, we illustrate how it can be generalized by employing standard approximations, namely SCE and exact-exchange (EXX).~\cite{ShaHor-PR-53, TalSha-PRA-76} 





The \textsl{conditional} amplitude, $\Phi(\sigma, 2,\dots, N ; \rr)$, defined in the framework of the EEF is
\begin{equation}\label{eq:CAdef}
\Phi(\sigma, 2,\dots, N ; \rr)\coloneqq\frac{\Psi (1, \dots, N)\, \sqrt{N}}{\sqrt{\ron (\rr)}}
\end{equation}
where $\Psi$ and $\ron$ are respectively the ground state (GS) wavefunction and GS density of the $N$-electron Hamiltonian under study and $i=\sigma_i \rr_i $ indicates the spatial and spin coordinates of the $i$-th electron. The subscript $1$ from the coordinates of electron $1$, chosen as reference is dropped, i.e. $\sigma_1=\sigma$ and $\rr_1=\rr$.

The Hartree-XC potential, $v_\text{Hxc}(\rr)$, can be exactly decomposed as~\cite{BuiBaeSni-PRA-89}
\begin{align}\label{eq:KSpotdec}
v_\text{Hxc}(\rr) ={}& v_{\kin}(\rr) - v_{s,\kin}(\rr) \notag \\
&{}+ v_{N-1}(\rr) - v_{s,N-1}(\rr) + v_{\cond} (\rr)
\end{align}
in terms of the kinetic potential $v_{\kin}(\rr) \coloneqq \frac{1}{2}\int \abs{\nabla_{\rr}\Phi}^2\, \ud \sigma \ud 2 \dotsi \ud N$, 
the $N-1$ potential $v_{N-1}(\rr) \coloneqq \int \Phi^* \hat{H}^{N-1}\Phi\, \ud \sigma \ud 2 \dotsi \ud N -E_0^{N-1}$ and the conditional potential $v_{\cond}(\rr) \coloneqq \int \sum_{i=2}^N v_{ee}(\rr -\rr_i)\abs{\Phi}^2 \, \ud \sigma \ud 2 \dotsi \ud N$. We used the symbols 
$ \hat{H}^{N-1}$ and $E_0^{N-1}$ respectively for the Hamiltonian and the GS energy of the ionized system, and
$v_{ee}$ for any interaction function. Finally, $v_{s,\kin}$ and $v_{s,N-1}$ for the KS system are formally defined as for the interacting system, replacing $\Phi$ with the KS \textsl{conditional} amplitude obtained from~\eqref{eq:CAdef} when substituting $\Psi$ with the KS wavefunction.
Note that, in the case of two-electron singlets, the KS kinetic and $N-1$ potentials are zero.~\cite{GiaNeuBaeGie-JCTC-22}


In this work, we focus on the approximation of $v_\kin$ and $v_\nm1$ in the case of two-electron singlets.
Two-electron singlets are an excellent proxy for studying the features of the KS potential in the breaking of a chemical bond. 
This is essentially because the peaks and the steps structure arising specifically in dissociation processes are located in very low-density regions, where the overall density is dominated by the density of the sole highest-occupied molecular orbital (HOMO), occupied by the two electrons involved in the bond that is breaking. Therefore, we can regard these features as coming from a two-electron singlet system where only $v_\kin$ and $v_\nm1$ play a role. 
In turn, these are also the components of the total potential that have proven the most delicate to model since they depend on the interacting \textsl{conditional} amplitude, as opposed to $v_{s,\kin}$ and $v_{s, \nm1}$ that can be written explicitly in terms of the occupied KS orbitals. 

For two-electron singlets, the \textsl{conditional} density is simply given by the modulus squared of the \textsl{conditional} amplitude
\begin{equation}
\ron_\cond(\rr,\rr')= |\Phi (\rr';\rr)|^2
\end{equation}
and, as a result, one can conveniently rewrite in terms of $\ron_\cond$ the kinetic potential as
\begin{align}\label{eq:vkincond}
v_\kin(\rr) =\frac{1}{8} \int \frac{\left( \nabla_\rr \ron_\cond(\rr,\rr') \right)^2}{\ron_\cond (\rr, \rr')} \, \md \rr'
\end{align}
and the $N-1$ potential as
\begin{align}\label{eq:vnm1cond}
v_\nm1(\rr)=&\int \left(\frac{\left( \nabla_{\rr'} \ron_\cond(\rr, \rr')\right)^2}{8\, \ron_\cond (\rr, \rr')} + v(\rr')\,\ron_\cond(\rr, \rr') \right) \md \rr' \notag \\
& - E_0^{N-1}
\end{align}
where $v$ is the external potential of the problem under study.
Importantly, the \textsl{conditional} potential can always be written in terms of the \textsl{conditional} density
\begin{equation}\label{eq:vcondcond}
v_\cond(\rr) =\int \ron_\cond(\rr,\rr') v_{ee}(\rr -\rr') \ud \rr'
\end{equation}
In conclusion, we can determine each individual contributions to the total potential once we have a model for the \textsl{conditional} density.

To approximate this latter quantity, we propose the following \textit{ansatz}
\begin{equation}\label{eq:ncondansatz}
\ron_\cond^{f_i}(\rr,\rr')=\ron(\rr')\sum_{i=1}^{N-1} G\left(\rr' - f_i(\rr) \right)
\end{equation}
where $f_i$ are special functions to be discussed and $G (\rr)=e^{- \beta \, \rr^2}$ is a Gaussian function. Its width, $\beta$, is to be determined from the condition 
$\int \ron_\cond(\rr,\rr')\,\ud \rr'=1 \,\, \forall \,\, \rr$, which ensures that the probability of finding any of the $N-1$ electrons over all space is one regardless of where the reference electron is located. Note that expression \eqref{eq:ncondansatz} is valid for general number of electrons $N$, although in the examples we consider for testing, we limit ourselves to the case $N=2$.

The \textit{ansatz} of Eq.~\eqref{eq:ncondansatz} is inspired to the structure of the SCE conditional density, \cite{MirSeiGor-JCTC-12, PedCheWhiBur-PRB-22}
\begin{equation}\label{eq:ncondSCE}
\ron_\cond^\sce (\rr, \rr') =\sum_{i=2}^N
\delta(\rr'-f_{i,\sce}(\rr))
\end{equation}
where the $f_{i,\sce}$ are the infamous co-motion functions -- or just co-motions -- of the SCE limit. They give the classical positions of the $N-1$ electrons as a function of the position of a reference electron. 
Two formal differences distinguish the SCE \textsl{conditional} density [Eq.~\eqref{eq:ncondSCE}] from our \textit{ansatz} in Eq.~\eqref{eq:ncondansatz}, namely: (i) the Dirac delta has been fattened into a Gaussian function and (ii) the Gaussian is furthermore weighted by $n(\rr')$.
Both these changes reduce the excessive correlation among electrons present in the SCE description. Correction (i) allows for quantum fluctuations around the parametric curve given by the co-motions.  Such an approach has appeared in different studies that moved from the leading order in the strong-interaction limit to the next.~\cite{GorVigSei-JCTC-09, SeiGiaVucFabGor-JCP-18, GerGroGor-JCTC-19}
 Secondly, via correction (ii) the probability distribution of any of the $N-1$ electrons does not depend solely on the position $\rr$ of the reference electron but also on the probability of having any other electron in $\rr'$.
Consequently, the term $\sum_{i=1}^{N-1} G\left(\rr' - f_i(\rr) \right)$ is recognized as an approximate pair-correlation function.

In addition to (i) and (ii), the shape of the co-motions itself is another factor that greatly influences the structure of the approximate \textsl{conditional} density.
The concept of the co-motions has been introduced in electronic structure by the SCE framework and has remained relatively limited to this extreme scenario. Here, we explore the possibility of determining effective co-motions that can give a good description of the relative position of electrons at the physical regime.  To distinguish them from the other effective models that will be proposed, we label the usual co-motions, associated to the SCE limit, as $f_{i,\sce}$. The symbol $f_i$ in Eq.~\eqref{eq:ncondansatz} encompasses any function that plays the role of a co-motion: determining the position of any of the $N-1$ electrons as a function of the reference one.

Our approach is somewhat reminiscent of the construction of the multi-radii functional (MRF).~\cite{VucGor-JPCL-17,VucBah-JCTC-23}
There, one constructs an approximate spherically-averaged \textsl{conditional} density from effective radial distances among electrons.
Here, instead, we consider not only the radial distances, but effective co-motions
where the information on the geometric orientation of electron pairs within the molecular structure is retained, aiming to model the full Hartree-XC potential, not (only) the energy density as in MRF. 

To test our \textit{ansatz} and the approximate potentials that follow, we adopt a simple $1$D model system that is paradigmatic of a stretched heteronuclear molecule and for which extremely accurate data are available.~\cite{GiaNeuBaeGie-JCTC-22}
The Hamiltonian considered is
\begin{align}\label{eq:genHam}
\hat{H} =\sum_{i=1,2} \left(- \frac{1}{2}\frac{\partial^2}{\partial x_i^2} + v(x_i)\right)+ \frac{1}{\sqrt{c+(x_1 -x_2)^2}}
\end{align}
with 
$
v(x) = - \frac{1}{\sqrt{a + \left( x - \frac{R}{2}\right)^2}}-\frac{1}{\sqrt{b + \left( x + \frac{R}{2}\right)^2}}.
$
The repulsive interaction potential 
and the external potential
are soft-Coulomb interaction functions where the Coulomb singularity in $1$D is softened by a parameter at the denominator. The external potential, $v$, features two wells representing the (different) nuclei, separated by a distance $R$. Nucleus $A$ is located at $x=\frac{R}{2}$, with potential value $1/\sqrt{a}$, and nucleus $B$ is at $x=-\frac{R}{2}$, with potential value $1/\sqrt{b}$. For the softening parameters, we have chosen $a=0.7\,a_0$, $b=2.25\,a_0$  and $c=0.6\,a_0$. With such parameters, the ionization potentials of the isolated fragments, $\Delta I$, is $0.3\,E_h$, reproducing the $\Delta I$ of the real LiH molecule.\cite{TemMarMai-JCTC-09, GiaNeuBaeGie-JCTC-22}
 The exact conditional amplitude $\Phi(x_2; x_1)$ and corresponding kinetic, $N-1$ and conditional potentials 
 have been calculated for this model at various internuclear distances.~\cite{GiaNeuBaeGie-JCTC-22}

The procedure to approximate the components of the Hartree-XC potential, including the peaks and the steps features, is as follows:
\begin{enumerate} 
\item defining an effective distance, $D(x)$, between the two particles as a function of the position of the reference electron;
\item determining the co-motions in terms of $D(x)$;
\item constructing the \textsl{conditional} density using the co-motions obtained at point 2. from Eq.~(\ref{eq:ncondansatz});
\item using the \textsl{conditional} density to calculate the kinetic [Eq.~(\ref{eq:vkincond})], $N-1$ [Eq.~(\ref{eq:vnm1cond})] and \textsl{conditional} [Eq.~(\ref{eq:vcondcond})] potentials.
\end{enumerate} 
To initiate our procedure, thus for point 1., we take as a measure of the distance between the electron pair the inverse of the \textsl{conditional} potential, as done in the MRF approximation,
\begin{equation}\label{eq:Dofx}
D(x)= \frac{1}{v_\cond(x)}
\end{equation}
essentially regarding the two particles as point charges. In Eq.~(\ref{eq:Dofx}), in principle any $v_\cond$ can be inserted, since it is only used at this stage of the procedure to determine the co-motions and, in turn, the \textsl{conditional} density. To demonstrate how this strategy can effectively improve the \textsl{conditional} potential, we begin by using the exact \textsl{conditional} potential and later we shall consider approximate \textsl{conditional} potentials. Note that, because of the soft-Coulomb interaction in our model, the distance is more precisely
$D(x)=\sqrt{\frac{1-c\, v_\cond(x) ^2}{v_\cond(x) ^2}}$
However, this is virtually indistinguishable from Eq.~\eqref{eq:Dofx}
for large $R$ (we take $R=$\SI{11}{\bohr} as representative of large $R$).

\begin{figure}
	\includegraphics[width=0.9\linewidth]{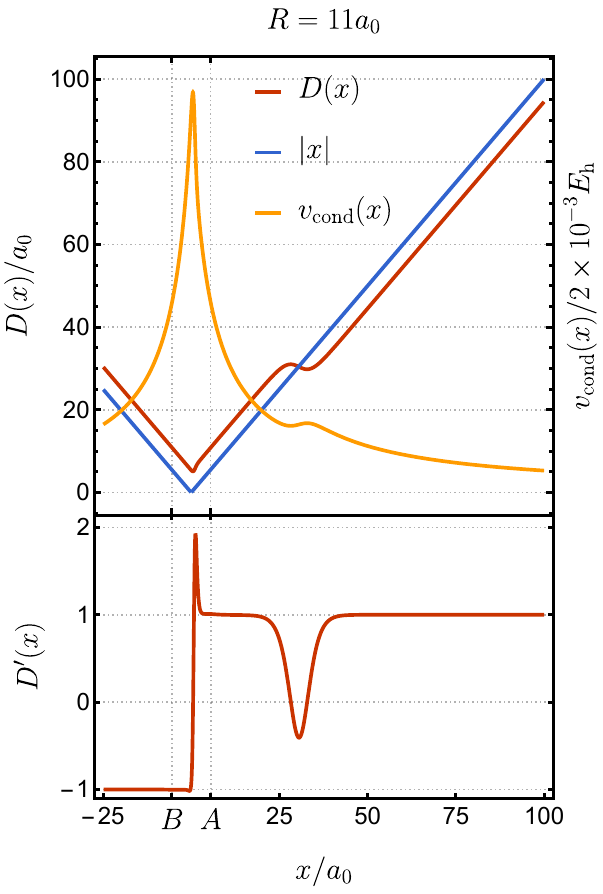}
	\caption{Upper panel: classical distance between particles, $D(x)$  [Eq.~\eqref{eq:Dofx}] as a function of the position of the reference electron $x$. The functions $|x|$ and $v_\cond(x)$ are also plotted for comparison. This latter is enhanced by a factor 500, i.e. in units of $2\times 10^{-3}\,E_h$. Lower panel: derivative of the distance between particles, $D'(x)$. The two transition regions where the derivative is not constant corresponds to the two maxima of the conditional potential.}
	\label{fig:DDpofx}
\end{figure}

In Fig.~\ref{fig:DDpofx} (upper panel), we plot $D(x)$ of Eq.~(\ref{eq:Dofx}) using the exact \textsl{conditional} potential, together with $|x|$ (for reference) and the \textsl{conditional} potential. $D(x)$ is decreasing as $-x$ (modulo a constant) roughly up to the global maximum of the \textsl{conditional} potential, past which it starts increasing as $x$.
One remarkable feature of $D(x)$ is the switch from above to below $x$ around \SI{31}\bohr , where the repulsion between electrons is locally increasing. This overall behaviour is fully appreciated in the derivative $D'(x)$, shown in the lower panel. 

The local maximum (shoulder) in the \textsl{conditional} potential in addition to the global one around the bond midpoint was first identified in Ref.~\citenum{GiaNeuBaeGie-JCTC-22}. Both maxima are related to positions of the reference electron for which the \textsl{conditional} amplitude changes from resembling the ground state of the ion to resembling its first excited state or vice versa. These positions also relate to the two peaks of the kinetic potential $v_\kin$ and to the extent of the step structure in the $N-1$ potential $v_\nm1$. The reader is referred to Ref.~\citenum{GiaNeuBaeGie-JCTC-22} for an extensive discussion of these features of the exact potentials.
%
%

\begin{figure}
	\includegraphics[width=\linewidth]{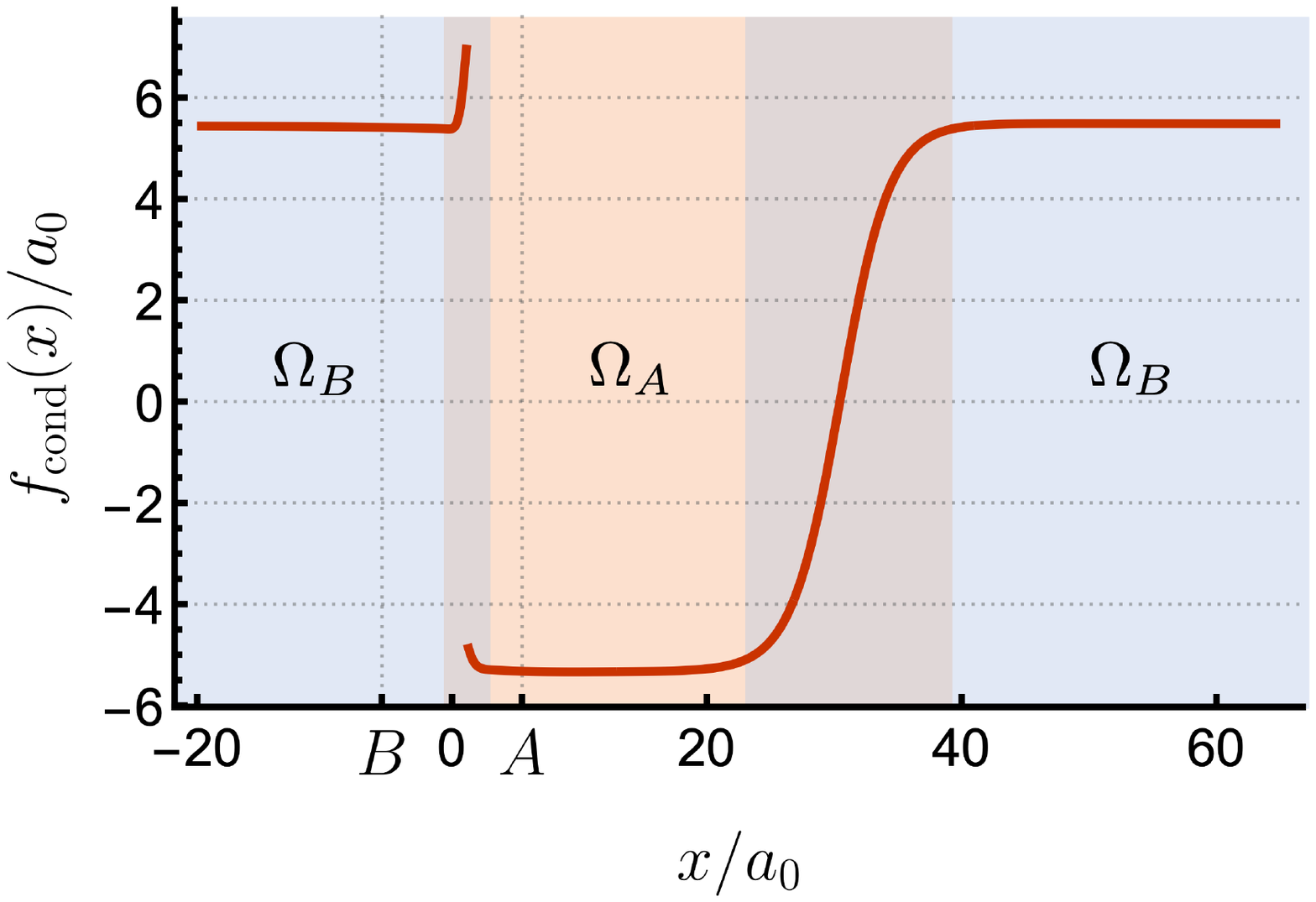}
	\caption{Effective co-motion, $f_\cond(x)$, giving the position of the remaining electron as a function of the position of the reference one, for internuclear distance $R=$\SI{11}{\bohr}. 
}
	\label{fig:fcond}
\end{figure}

We move to point 2., and we determine an effective co-motion to use in Eq.~\eqref{eq:ncondansatz}, for which we use the symbol $f_\cond(x)$ to indicate that it is the one obtained by using the exact $v_\cond$ in Eq.~(\ref{eq:Dofx}). 
Recalling that the co-motion is the classical position of an electron as function of the reference one in $x$, and thanks to the reduced dimensionality of the model, the distance between the electron pair is $D(x)= |x - f_\cond(x) |$, or
\begin{equation}\label{eq:ftilde}
f_\cond(x) = \begin{cases} x + D(x)  \quad x \leq x_\text{crit}\\x - D(x)\quad x > x_\text{crit}
\end{cases}
\end{equation}
where $x_\text{crit}$ is a critical value of $x$ to be assigned. To assign it, we first take a look at what is happening for large positive $x$. When $D(x)<x$, both electrons are to the right of the bond midpoint: the reference electron is being removed ($x \to \infty$) and we expect the remaining electron to sit around nucleus A at $\frac{R}{2}$, which is more attractive. As the reference electron is moved in the direction of the bond midpoint (towards negative values of $x$), 
the distance between the two electrons is decreased, until at some point their repulsion becomes dominant and the remaining electron is forced to move around nucleus B at $-\frac{R}{2}$, to the left of the bond midpoint. This is done smoothly, with the distance $D(x)$ gradually becoming greater than $x$ and going through an inflection point around \SI{31}\bohr . The sign of $ x - f_\cond(x) $ has remained so far positive while varying $x$. 
However, as the reference electron is moved closer and closer to the bond midpoint, the repulsion increases again until the other electron is forced to move back around nucleus A. Differently than the large positive $x$ scenario, though, in this case $ x - f_\cond(x) $ changes sign (at $x=x_\text{crit}$). 
By analogy with what happens in the large positive $x$ case, we pick as $x_\text{crit}$ the other inflection point of $D(x)$, i.e. where $D'(x)$ has its maximum in the lower panel of Fig.~\ref{fig:DDpofx}.
The result is a function that is discontinuous at $x=x_\text{crit}$ (see Fig.~\ref{fig:fcond}).

Figure~\ref{fig:fcond} shows the co-motion $f_\cond(x)$ obtained as described above. The physical interpretation of this co-motion is quite simple and reasonable, yielding the position of the remaining electron as function of the position $x$ of the reference electron. In Fig.~\ref{fig:fcond}, we label $\Omega_A$ the region where the reference electron feels only the field of nucleus A and vice versa for $\Omega_B$.~\cite{GiaNeuBaeGie-JCTC-22} In $\Omega_A$ (shaded red area), when the reference electron is either sitting on nucleus A or on its right, up to $\approx$\SI{20}{\bohr}, the position of the remaining electron is constant and negative, $\approx$\SI{-5.35}{\bohr}, thus it is located around nucleus B. In turn, when the reference electron is on nucleus B or on its left, the position of the remaining electron is constant and positive, $\approx$\SI{5.35}{\bohr}, thus it is located around nucleus B, as indicated in Fig.~\ref{fig:fcond} in $\Omega_B$ (shaded blue areas). When the reference electron is being ionized and is located at large positive values of $x$, the remaining electron is close to nucleus A, thus the reference electron effectively feels the field of nucleus B (therefore, the region on the right is labelled $\Omega_B$). The transition regions (in purple) correspond to regions where the reference electron feels both nuclei.

%
%
As indicated in point 3., we then construct an approximate \textsl{conditional} density via Eq.~\eqref{eq:ncondansatz}. The width of the Gaussian, $\beta^{f_\cond}$, associated to the co-motion $f_\cond$, is determined numerically from the normalization condition for the \textsl{conditional} density. The function $\beta^{f_\cond}(x)$ is plotted in Fig.~(SM1) of the Supplemental Material, together with the widths $\beta^{f_\sce}(x)$ and $\beta^{f_\text{EXX}}(x)$, which are associated to different kinds of co-motions (SCE and EXX) and will be illustrated later. 

\begin{figure*}
	\includegraphics[width=.8\linewidth]{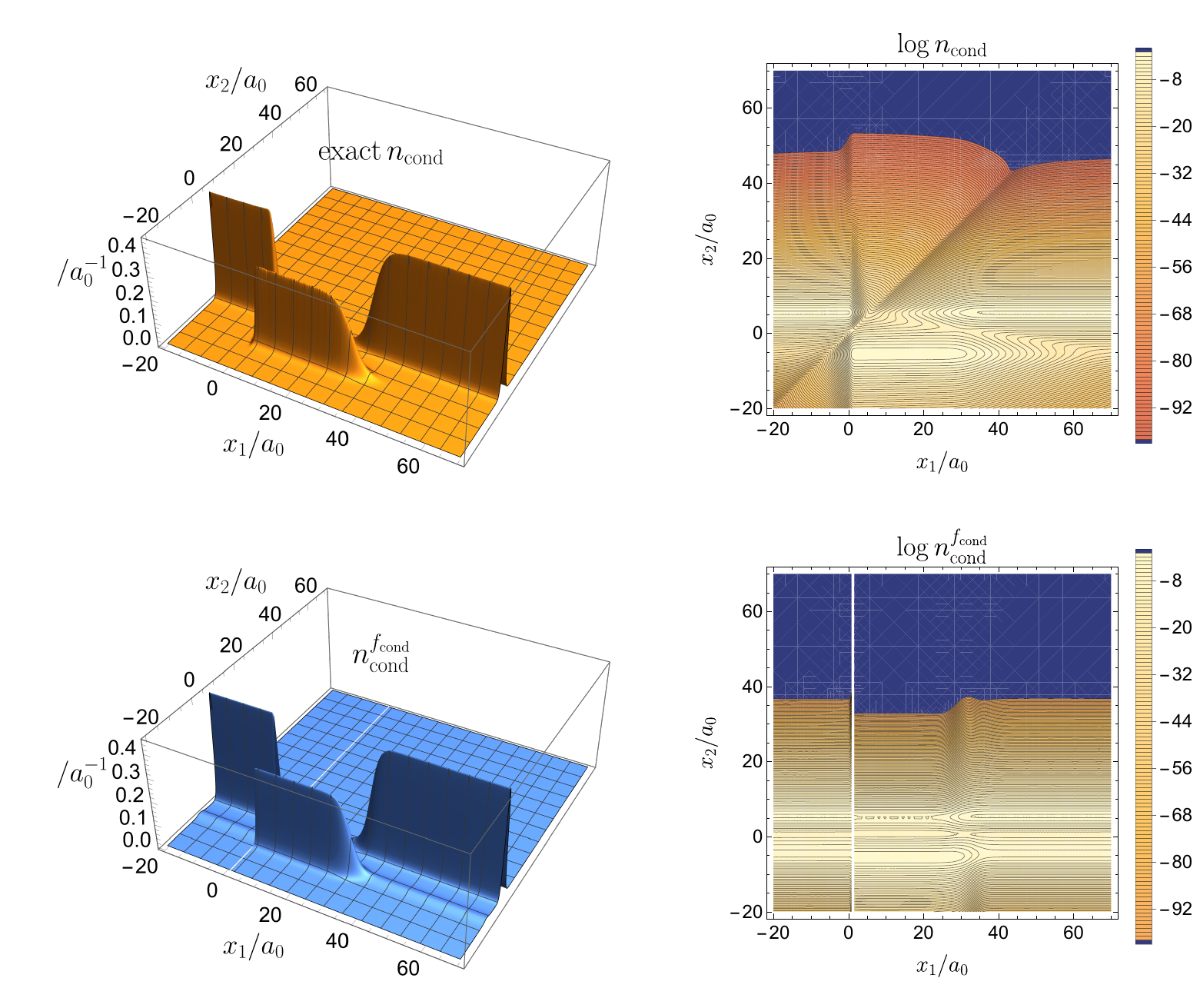}
	\caption{Upper panels: exact \textsl{conditional} density $n_\cond (x_1, x_2)$ (left) and contour plot of its logarithm (right). Lower panels: approximate \textsl{conditional} density $n_\cond^{f_\cond} (x_1, x_2)$ (left) and its logarithm (right).}
	\label{fig:ncondfig}
\end{figure*}

Figure~\ref{fig:ncondfig} is the first main result of this work: we show our approximate \textsl{conditional} density $n_\cond^{f_\cond} (x_1, x_2)$, obtained from the \textit{ansatz}~(\ref{eq:ncondansatz}), and we compare it with the exact one. In the left column of the figure, where the \textsl{conditional} densities are plotted as $3$D surfaces,  the accuracy of the proposed \textit{ansatz} stands out. The only region of appreciable deviation is around $x_2\approx -\frac{R}{2}$ and $x_1 \in \Omega_B$, where there is an excess of probability density (as well as around $x_2 \approx \frac{R}{2}$ and $x_1 \in \Omega_A$, less visible in the figure). Plotting the natural logarithm of both \textsl{conditional} densities stresses other structural differences between the two, notably their different decay along the diagonal $x_1=x_2$.

\begin{figure}
	\includegraphics[width=\linewidth]{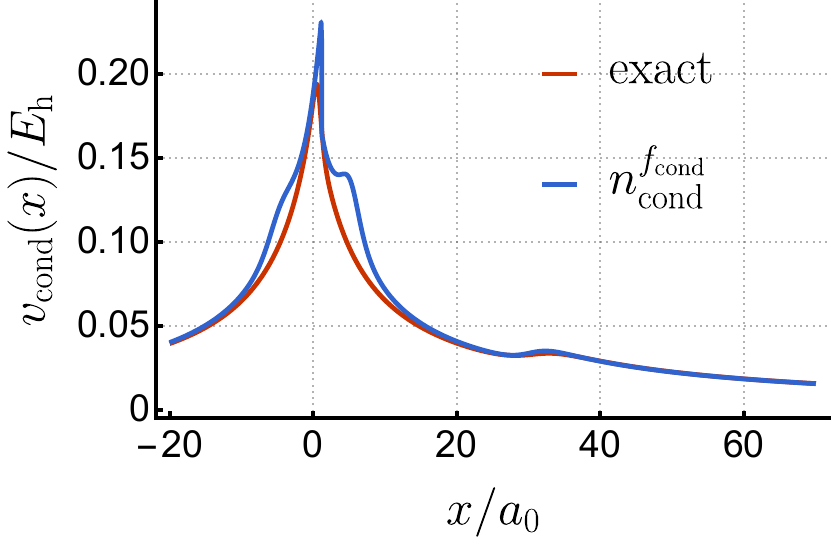}
	\includegraphics[width=\linewidth]{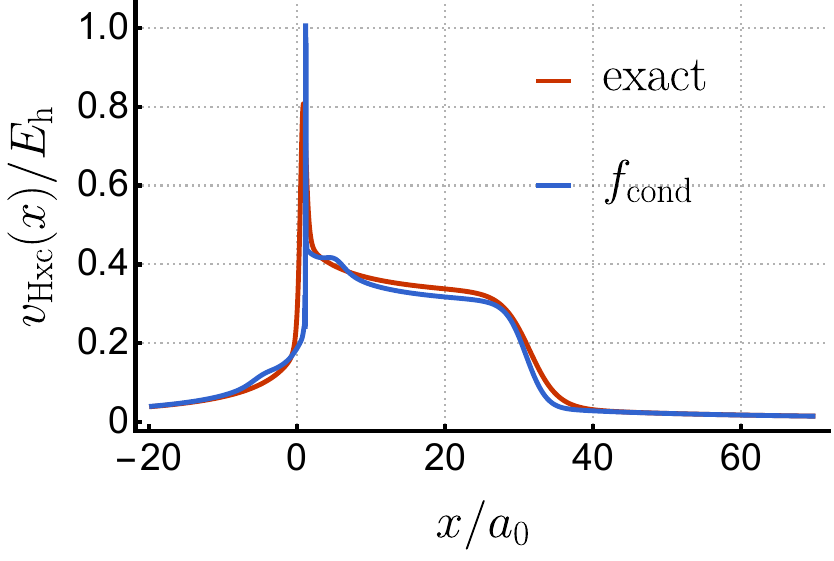}
	\caption{Results obtained from $f_\cond$ and $n_\cond^{f_\cond}$. Upper panel: Approximate \textsl{conditional} potential (blue) in comparison with the exact one (red). Lower panel: Approximate Hartree-XC potential $v_\text{Hxc}^{f_\cond}$ (blue) in comparison with the exact one (red).
}
	\label{fig:vcond and vhxc fcond} 
\end{figure}
Finally, according to point 4., we can construct the local potentials of Eqs.~\eqref{eq:vkincond}, \eqref{eq:vnm1cond} and \eqref{eq:vcondcond}, to assess how these deviations of the approximate \textsl{conditional} density impact the approximate potentials. We begin by analyzing the approximate \textsl{conditional} potential in comparison with the exact, which was the source of our effective co-motion. Both are plotted in Fig.~\ref{fig:vcond and vhxc fcond} (upper panel). As visible, the approximate potential (in blue) shows excessive repulsion around $\pm \frac{R}{2}$, as a consequence of the excess of probability density discussed in connection with Fig.~\ref{fig:ncondfig}. Nonetheless, the overall shape of the potential appears quite accurate: especially accurate is the local maximum around \SI{31}{\bohr}.
Figure~\ref{fig:vcond and vhxc fcond} (lower panel) is the second main result of this work. We show the sum of all the potentials $v_\cond + v_\kin+v_\nm1$, which corresponds to an approximate Hartree-XC potential in two-electron singlets (in blue). The accuracy of the model proposed is quite remarkable, particularly in the step structure, which has been modelled satisfactorily without any resort to virtual KS orbitals. To the best of our knowledge, previous models of this structure insisted on the necessity for using virtual KS orbital(s) as input.~\cite{GriBae-IJQC-06, HelGou-JCTC-19} Here, instead, we demonstrate that it is possible to reproduce this peculiar feature within an orbital-independent framework. Because the $v_\cond$ that has been used to initiate the whole procedure was the exact one, this result is a proof-of-principle. However, we will now repeat a similar construction using approximate forms of $v_\cond$.


First, let us use the SCE co-motion, $f_\sce$. 
To avoid the logarithmic divergence of $f_\sce$ at the point where the density integrates to one (which is the $x_\text{crit}$ of Eq.~\eqref{eq:ftilde} in the SCE co-motion), we adopt the model co-motion developed in Ref.~\citenum{GiaGor-JPCA-20}, which is based on the analytical density 
$\ron_\text{mod} = \ron_A + \ron_B$ with $\ron_{A (B)} = \frac{\mathfrak{a} (\mathfrak{b})}{2} e^{-\mathfrak{a} (\mathfrak{b}) | x - (+) \frac{R}{2} |}$. As seen from Fig.~(SM2) of the Supplemental Material, this model gives a reasonably accurate approximation of the exact SCE \textsl{conditional} potential using as coefficients the ones obtained from $\mathfrak{a}=2\sqrt{2 \epsilon_A} $ and $\mathfrak{b}=2\sqrt{2 \epsilon_B} $ and $\epsilon_A= -0.867$ 
and  $\epsilon_B =-0.568 $ the two lowest eigenvalues of the ionized system  $\hat{h}=-\frac{\partial^2 }{2 \, \partial \, x^2} + v (x)$. The corresponding  width, $\beta^{f_\sce}$, to construct our approximate \textsl{conditional} density, $n_\cond^{f_\sce}$, is shown in blue in Fig.~(SM1) of the Supplemetal Material. 
\begin{figure*}
	\includegraphics[width=.48\linewidth]{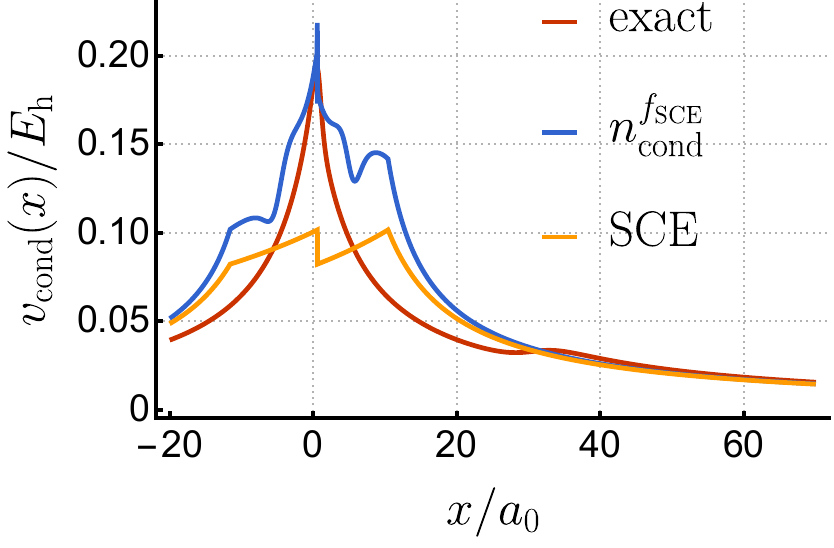}
	\includegraphics[width=.48\linewidth]{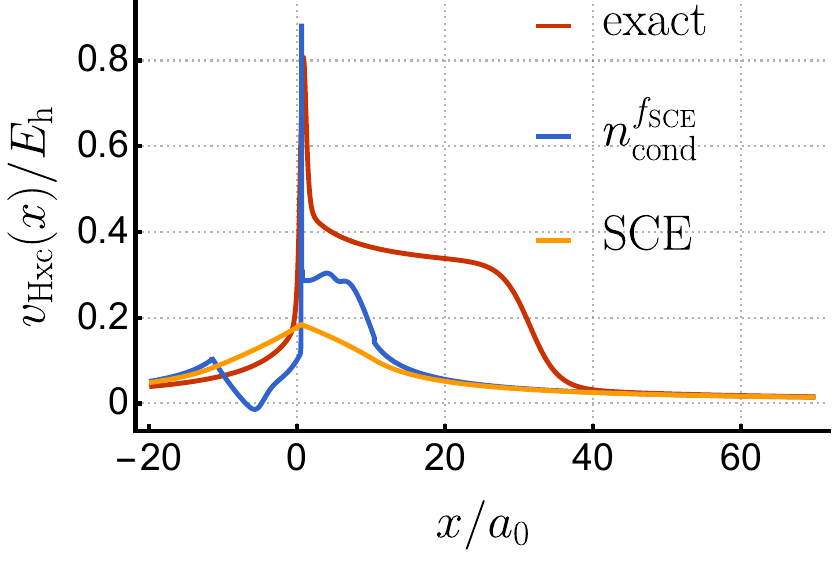}
	\includegraphics[width=0.48\linewidth]{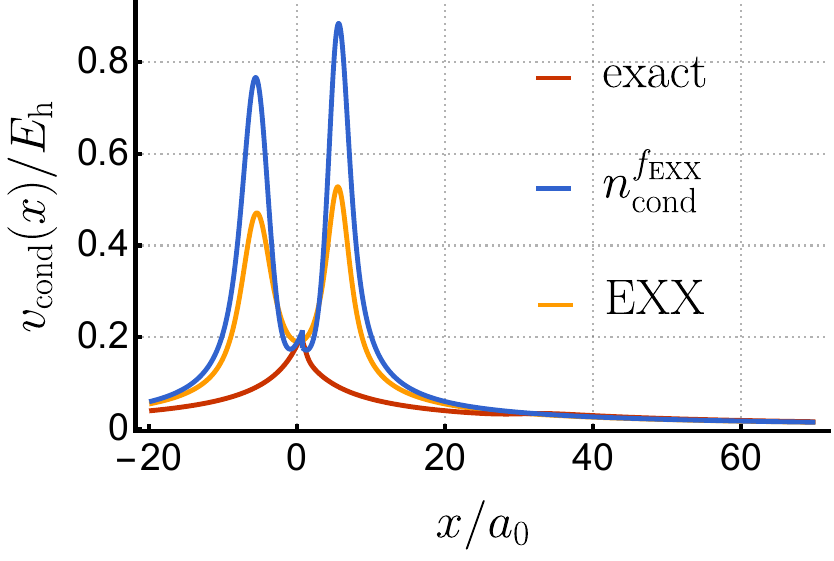}
	\includegraphics[width=0.48\linewidth]{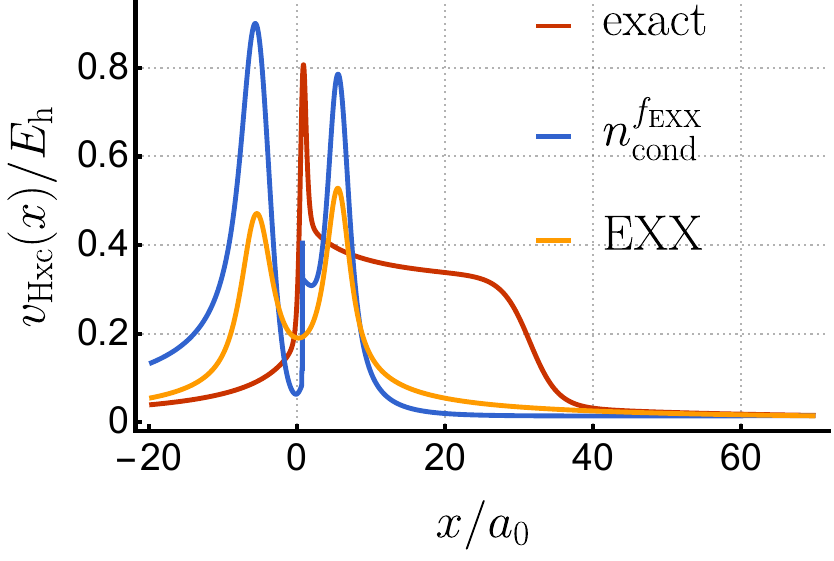}
	\caption{Upper panels: Results obtained from $f_\sce$ and $n_\cond^{f_\sce}$. Left: Approximate \textsl{conditional} potential coming from $n_\cond^{f_\sce}$ (blue) in comparison with the exact one (red) and with the SCE \textsl{conditional} potential (yellow). Right: Approximate Hartree-XC potential coming from $n_\cond^{f_\sce}$ (blue) in comparison with the exact one (red) and with the SCE Hartree-XC potential (yellow). Lower panels: Results obtained from $f_\text{EXX}$ and $n_\cond^{f_\text{EXX}}$. Left: Approximate \textsl{conditional} potential coming from $n_\cond^{f_\text{EXX}}$ (blue) in comparison with the exact one (red) and with the EXX \textsl{conditional} potential (yellow). Right: Approximate Hartree-XC potential coming from $n_\cond^{f_\text{EXX}}$ (blue) in comparison with the exact one (red) and with the EXX Hartree-XC potential (yellow).
}
	\label{fig:vcond and vhxc fsce and fexx}
\end{figure*}

In Fig.~\ref{fig:vcond and vhxc fsce and fexx} (upper left panel), the resulting \textsl{conditional} potential (blue) is shown in comparison with the exact (red) and the SCE (yellow) \textsl{conditional} potentials. Although it displays some incorrect features and excessive repulsion, it appears qualitatively much closer to the shape of the exact potential, $v_\cond$, than to the shape of $v_\cond^\sce$. This result validates the strategy proposed here, aiming to modify the structure of the SCE \textsl{conditional} density [Eq.\eqref{eq:ncondSCE}] as per the \textit{ansatz}~\eqref{eq:ncondansatz}, 
in order to capture more features of the electron-electron interaction in the physical regime.
The total Hartree-XC potential coming from $n_\cond^{f_\sce}$ (blue) is shown in Figure~\ref{fig:vcond and vhxc fsce and fexx} (upper right panel) in comparison with the exact (red) as well as the SCE (yellow) Hartree-XC potential.
\cite{GiaVucGor-JCTC-18, GiaGor-JPCA-20} Also for the total potential, it is remarkable that the approximate potential from our \textit{ansatz} more closely resembles the exact potential rather than the SCE potential, to which the specific co-motion belongs.


Second, we apply our strategy to the exact-exchange approximation\cite{ShaHor-PR-53, TalSha-PRA-76} (EXX). 

As described in the Supplemental Material, %
the \textsl{conditional} potential and the total Hartree-XC potential in the EXX approximation for two-electron singlets coincide, i.e. $v_\cond^\text{EXX}=v_\text{Hxc}^\text{EXX}$.

%
%
To construct an effective co-motion from $v_\cond^\text{EXX}$, we first extract an effective distance as in Eq.~\eqref{eq:Dofx}.  Then, we choose as  $x_\text{crit}$ the $x$ value where the density within the two nuclei is minimal, as this point is typically very close to the point where the density integrates to one ($x_\text{crit}$ for $f^\sce$) as well as to the location of the maximum of the kinetic potential ($x_\text{crit}$ for $f^\cond$).\cite{HodRamGod-PRB-16, GiaVucGor-JCTC-18} The associated width, $\beta^{f_\text{EXX}}$, is shown in yellow in Fig.~(SM1) of the Supplemental Material. The resulting \textsl{conditional} potential (blue) is shown in Figure~\ref{fig:vcond and vhxc fsce and fexx} (lower left panel) together with the exact (red) and the EXX (yellow) conditional potentials. Although our approximation increases the repulsion around the positions of the nuclei if compared to the source potential ($v_\cond^\text{EXX}$), it does introduce a qualitative correct feature: a small peak around the bond midpoint. This is a direct consequence of the structure of the proposed \textit{ansatz}: the use of an effective co-motion introduces some correlation effects that are absent from the source potential. The same effect is visible in the total Hartree-XC potential, which now features a hint of step structure completely absent from the EXX potential.

Unfortunately, neither the SCE- nor the EXX-modified potentials model the accurate extent of the step structure, because their source potentials do not feature the local maximum of the exact \textsl{conditional} potential. Therefore, their
 effective co-motions do not model the transition around \SI{31}{\bohr} from one nucleus to the other, differently than $f_\cond$ (Figure~\ref{fig:fcond}). 
 
In summary, in this work, we have presented a strategy to model non-local features of the exact Hartree-XC potential that brings together the exact electron factorization and the strictly-correlated-electron limit of DFT. The features targeted are crucial to describe the breaking of a chemical bond. We have provided a proof-of-principle demonstration that it is possible to capture such non-local features of the Hartree-XC potential, particularly its step structure, without resorting to any KS virtual orbitals, which seems to represent a shift compared to previous common understanding.
We have tested our strategy on a system that is paradigmatic of the challenge faced by current KS-DFT approximations in modelling molecular dissociation. The key ingredients of the proposed procedure are the \textsl{conditional} potential and the \textsl{conditional} density. The exact \textsl{conditional} potential, along with the SCE and EXX \textsl{conditional} potentials, has been used as a source of an effective co-motion to approximate the \textsl{conditional} density. We have obtained excellent accuracy using the exact potential as input and important indications that our strategy is promising even with approximate starting ingredients, correctly reducing the perfect correlation in SCE  and bringing in some correlation in EXX (Fig.~\ref{fig:vcond and vhxc fsce and fexx}). Several aspects need to be further investigated and refined, building upon the fundamental concepts proposed in this work. For instance, the discontinuity obtained in the effective co-motions may be an artifact of the one-dimensional framework. Therefore, testing our strategy on simple spherical systems seems a natural next step. In general, models of more widely applicable co-motions need to be developed for our strategy to become suitable for systems of chemical interests. The concept of intracule density or the Overhauser model,~\cite{Ove-CJP-95} where the probability of having an electron pair separated by a distance $r_{12}$ is the main variable, may be exploited to derive effective co-motions in the future.




This project has received funding from the European Union’s Horizon 2020 research and innovation programme under the Marie Sk\l odowska-Curie grant agreement No 101104947


\bibliography{project0}

\clearpage
\section*{Supplemental Material}
\makeatletter 
\renewcommand{\thefigure}{SM\@arabic\c@figure}
\renewcommand{\thesection}{SM\@arabic\c@section}
\renewcommand{\theequation}{SM\@arabic\c@equation}
\makeatother
\setcounter{figure}{0} 
\setcounter{equation}{0} 
\setcounter{section}{0} 
\section{Additional figures}
\begin{figure}[h!]
	\includegraphics[width=.5\linewidth]{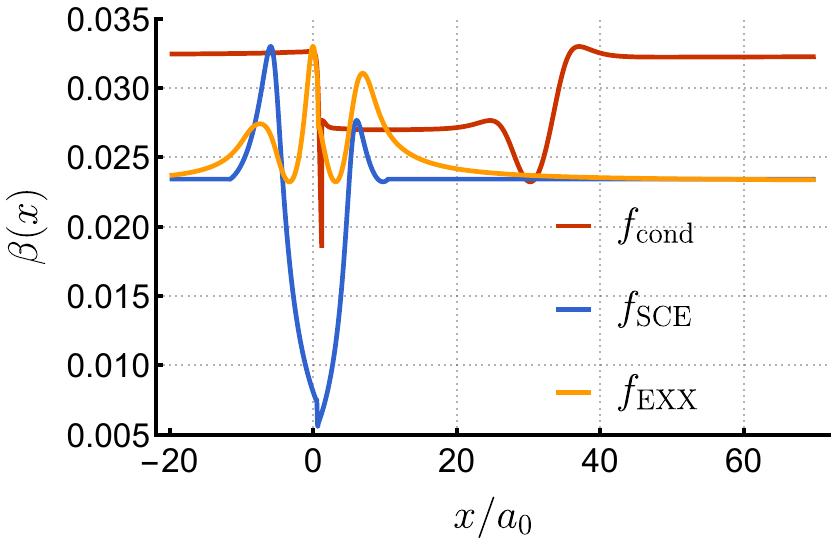}
	\caption{Width of the Gaussian function of Eq.~(7) of the main text associated to three different effective co-motions: $f_\cond$ (red), $f_\sce$ (blue) and $f_\text{EXX}$ (yellow).
}
	\label{fig:betas}
\end{figure}

\begin{figure}[h!]
	\includegraphics[width=0.5\linewidth]{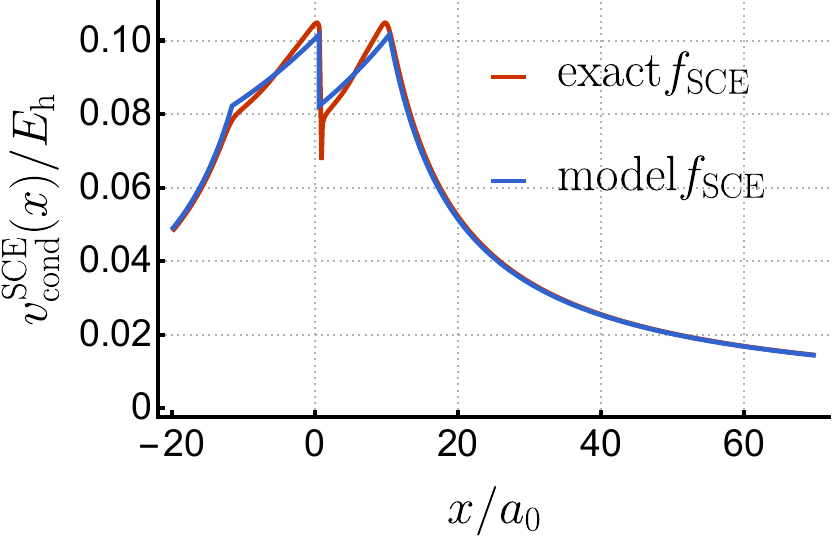}
	\caption{Approximate SCE conditional potential coming from the analytical density $\ron_\text{mod}$ (blue) in comparison with the exact (numerical) SCE conditional potential coming from the ground-state density of Eq.~(9) of the main text (red).
}
	\label{fig:SCEcond}
\end{figure}

\section{Exact-exchange potential in two-electron singlet systems}
In two-electron singlets, the exact-exchange approximation (EXX) simply corresponds to a total Hartree-XC potential equal to half of the Hartree potential, i.e.  $v_\text{Hxc}^\text{EXX}=v_\text{H}+v_\text{x} =\frac{v_\text{H}}{2} $, as the correlation contribution is always neglected in EXX while the exchange contribution merely corrects the self-interaction error in $v_\text{H}$ in this case. 
Because of that, $v_\text{x}$ also has no response component.
In general cases, one has
\begin{eqnarray}
& E_\text{x}[\ron] = \int \ron(\rr) v_\text{x-hole} (\rr) \ud \rr \label{eq:noresp} \\
& v_\text{x}(\rr) =\frac{\delta  \, E_\text{x}[\ron]}{\delta \, \ron (\rr)} = v_\text{x-hole} (\rr) + v_\text{x,resp} (\rr), \label{eq:respwork} 
\end{eqnarray}
where $v_\text{x-hole} $ is the energy density that yields back the exchange energy via Eq.~\eqref{eq:noresp}, while $v_\text{x,resp}$ is the response component, which is not directly involved in the energy formula. 
However, $v_\text{x,resp}$ is zero for two-electron singlets.\cite{GiaVucGor-JCTC-18} That is because $v_\text{x}$ inherits the functional form of the Hartree potential, which is purely an energy density as $U_\text{H} =\int \ron (\rr) v_\text{H} (\rr) \ud \rr$ and $\frac{\delta  \, U_\text{H}[\ron]}{\delta \, \ron (\rr)}=v_\text{H} (\rr) $.
(Because the exchange potential is defined modulo a constant, one can always shift the total exchange potential such that $E_\text{x} = \int v_\text{x}(\rr) \ron(\rr) \ud\rr$, but the fact that a component of the total exchange potential, which we call $ v_\text{x,resp}$, has no role in giving $E_\text{x}$, as per Eqs.~\eqref{eq:noresp} and \eqref{eq:respwork}, remains true regardless of the gauge. In all potentials considered in our work, we adopt the usual gauge in which the potential vanishes asymptotically.)

In turn, the conditional potential corresponds to the sum of the Hartree plus the XC-hole contribution,\cite{GiaVucGor-JCTC-18} $ v_\text{cond}(\rr) = v_\text{H}(\rr) + v_\text{xc-hole}(\rr)$,
where $v_\text{xc-hole}$ is the exchange-correlation energy density that yields back the total XC energy via $E_\text{xc}=\int v_\text{xc-hole}(\rr) \ron(\rr) \ud \rr$.
Then, in the EXX approximation for two-electron singlets, one concludes that
\begin{equation}
    v_\text{Hxc}^\text{EXX}(\rr)=v_\text{H}(\rr)+v_\text{x-hole}(\rr)=v_\cond^\text{EXX}(\rr).
\end{equation}

\end{document}